\documentclass{egpubl}
\usepackage{sca2020}
 
\SpecialIssuePaper         

\usepackage{amssymb}
\usepackage{amsmath}
\usepackage{float}
\usepackage{subcaption}
\usepackage{xcolor}
\usepackage[T1]{fontenc}
\usepackage{dfadobe}  
\usepackage{tipa}
\usepackage[utf8]{inputenc}

\usepackage{cite}  
\BibtexOrBiblatex
\electronicVersion
\PrintedOrElectronic

\ifpdf \usepackage[pdftex]{graphicx} \pdfcompresslevel=9
\else \usepackage[dvips]{graphicx} \fi

\usepackage{egweblnk} 

\title[Intuitive Facial Animation Editing based on a Generative RNN Framework]%
{Intuitive Facial Animation Editing Based On A Generative RNN Framework}

      \author[Elo\"{i}se Berson, Catherine Soladi\'{e} \& Nicolas Stoiber]
      {\parbox{\textwidth}{\centering Elo\"{i}se Berson,$^{1,2}$
        Catherine Soladi\'{e}$^{2}$
        and Nicolas Stoiber$^{1}$}
        \\
{\parbox{\textwidth}{\centering $^1$Dynamixyz, France\\
         $^2$CentraleSup\'{e}lec, CNRS, IETR, UMR 6164, F-35000, France}}}

\begin{document}

\teaser{
 \includegraphics[width=0.7\linewidth]{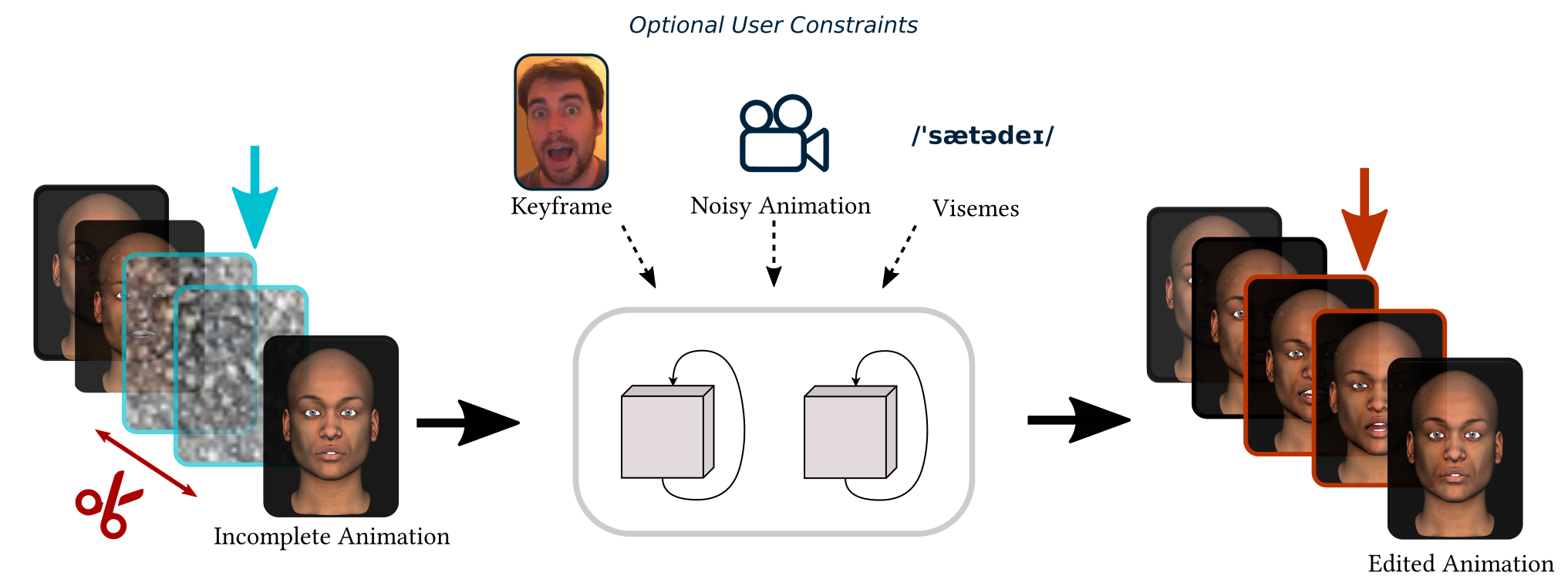}
    \centering
    \caption*{}
}

\maketitle
\begin{abstract}
For the last decades, the concern of producing convincing facial animation has garnered great
interest, that has only been accelerating with the recent explosion of 3D content in both entertainment and professional activities. 
The use of motion capture and retargeting has arguably become the dominant solution to address this demand.
Yet, despite high level of quality and automation performance-based animation pipelines still require
manual cleaning and editing to refine raw results, which is a time- and skill-demanding process.
In this paper, we look to leverage machine learning to make facial animation editing faster and more accessible to non-experts.
Inspired by recent \textit{image inpainting} methods, we design a generative recurrent neural network that generates realistic motion into designated segments of an existing facial animation, optionally following user-provided guiding constraints.
Our system handles different supervised or unsupervised editing scenarios such as motion filling during occlusions,
expression corrections, semantic content modifications, and noise filtering.
We demonstrate the usability of our system on several animation editing use cases.

\end{abstract}  

\section{Introduction}
Creating realistic facial animation has been a long-time challenge in the industry, historically relying on the craftmanship of few highly trained professional animators.
In the last three decades, the research community has produced methods and algorithms aiming to make quality facial animation generation accessible and widespread.
To this day, this remains a challenge due to the complexity of facial dynamics, triggering a plethora of spatiotemporal motion patterns ranging from subtle local deformations to large emotional expressions.
The emergence and increasing availability of motion capture (mocap) technologies have opened a new era, where realistic animation generation is more deterministic and repeatable.


The theoretical promise of mocap is the ability to completely and flawlessly capture and retarget a human performance, from emotion down the most subtle motion of facial skin.
In reality, even professional motion capture setups often fall short of a perfect animation result:
It is for instance usual that some part of a performance cannot be captured due to occlusions or unexpected poses.
For facial mocap, popular video-based technologies have known flaws as well: the camera resolution limits the capture precision, while signal noise, jitter, and inconsistent lighting can impair its robustness.
In addition to technical considerations, performance-based animation also lacks flexibility when the nature of captured motion does not match the desired animation result,
when the animation intent suddenly differs from what was captured, or the performer cannot or has not performed the requested motions.
Animation editing -or \textit{cleaning} as it is often called- is therefore unavoidable, and often the bottleneck of modern performance-based animation pipelines.

The editing task usually consists of selecting an unsatisfactory or corrupted time segment in the animation, and either correct or replace animation curves in that segment using computational or manual methods.
Several automatic motion completion systems have been developed based on simple interpolation between user-specified keyframes~\cite{parke_computer_1972}, usually with linear or cubic polynomials,
because of their simplicity and execution speed.
While interpolation has proven efficient for short segments with dense sets of keyframes, the smooth and monotonous motion patterns they produce are far from realistic facial dynamics when used on longer segments.
In most cases today, animation cleaning thus relies on \textit{keyframing}: having artists replace faulty animation with numerous carefully-crafted keyframes to interpolate a new animation.
Not only is \textit{keyframing} a time- and skill-demanding process, it requires acting on several of the character's low-level animation parameters, which is not intuitive for non-experts.

At the origin of this work is the parallel we draw between editing an animation and performing image inpainting.
Image inpainting aims at replacing unwanted/missing parts of an image with automatically generated pixel patterns, so that the edited image looks realistic.
In animation editing, we pursue the same objective, substituting 2D spatial pixel patterns for 1D temporal motion signals.
Inspired by recent advances in image inpainting frameworks, we present a machine-learning-based approach that makes facial animation editing faster and more accessible to non-experts.
Given missing, damaged, or unsatisfactory animation segments, our GAN-based system regenerates the animation segment, following few discrete semantic user-guidance such as keyframes, noisy signals, or a sequence of visemes.

Previous works have proposed to use a reduced set of low-dimensional parameters to simplify animation editing, either
based on temporal Poisson reconstruction from keyframes~\cite{akhter_bilinear_2012, seol_spacetime_2012} or based on regression from semantic-level temporal parameters~\cite{berson_robust_2019}.
While achieving smooth results, they require dense temporal specifications to edit long sequences.
In addition, in many cases where the capture process has failed to produce an animation (occlusion, camera malfunctions),
no input animation is available to guide the result; hence the animator has to create the whole missing sequence from scratch.
Our system handles all those cases by leveraging a GAN framework~\cite{goodfellow_generative_2014} to generate animation curves either from guidance inputs or unsupervised.
GANs have demonstrated impressive results at generating state-of-the-art results from little to no inputs in many tasks, such as image translation~\cite{isola_image--image_2017},
image inpainting~\cite{yu_free-form_2019, li_generative_2017}, and text-to-image~\cite{reed_generative_2016}.
Our system consists of a generator intending to create plausible sequences in designated segments in the input animation, and
a discriminator ensuring that the generated animation looks realistic.
To cope with learning the complex temporal dynamics of the facial motion, we design our generator as a bidirectional recurrent architecture, ensuring both past and future motion consistency.
Our system aims at providing an intuitive and flexible tool to edit animations. 
Hence, we provide the user with control over the edited animation through high-level guidance,
just as sketches enable semantic image manipulation in image inpainting scenarios~\cite{jo_sc-fegan_2019}.
Rather than specifying animation parameters, the user could guide the editing through semantic inputs,
such as sparse static expression constraints, visemes, or a noisy animation.
Our approach reduces both the time and the manual work currently required to perform facial animation editing, while retaining the flexibility and the creativity properties of the current tools. In summary, our primary contributions are:

\begin{itemize}
	\item A multifunctional framework that handles various high-level and semantic constraints to guide the editing process. 
          It can be applied to many editing use cases, such as long occlusions,
          expressions adding/changing, or viseme modifications.
	\item A generative and flexible system enabling fast unsupervised or supervised facial animation editing.
            Inspired by recent inpainting schemes, it leverages machine-learning-based signal reconstruction and transposes it in the facial animation domain.
            This framework allows editing motion segments of any length at any point in the animation timeline.

\end{itemize}

\section{Related Work}
In this paper, we propose a generative system for facial animation editing,
synthesizing new facial motions to fill missing or unwanted animation segments.
In this section, we point to relevant techniques for animation generation (Section~\ref{sec:facial_gene})
and motion editing (Section~\ref{sec:facial_edit}).
Finally, as our system can perform guided editing using semantic inputs,
such as keyframes or visemes, we review works related to facial reenactment (Section~\ref{sec:facial_reen}).

\subsection{Animation Generation}
\label{sec:facial_gene}
In this section, we discuss existing animation synthesis techniques that rely only on sparse or no explicit external constraints, encompassing methods that automatically generate motion transitions between keyframes (Section~\ref{subsec:keyframing_rw}), or techniques predicting motion sequences based on past context (Section~\ref{subsec:motion_pred_rw}).

\subsubsection{Keyframing-based Editing}
\label{subsec:keyframing_rw}
The most basic and widespread form of animation generation is keyframing.
Artists specify the configuration of character at certain key points in time and let an interpolation function generate the in-between motion.
Early works on facial editing focus on improving the keyframing process, providing an automatical solving strategy to map high-level static users’ constraints to the key animation parameters. 
User constraints are formulated as either 2D points such as image features~\cite{zhang_geometry-driven_2003}, motion markers~\cite{joshi_learning_2003}, strokes on a screen~\cite{dinev_user-guided_2018, cetinaslan_direct_2018, cetinaslan_sketch-based_2015} or the 2D projection of 3D vertices~\cite{zhang_spacetime_2004, chi_interactive_2017};
or 3D controllers like vertices position on the mesh~\cite{lewis_direct_2010, anjyo_practical_2012, tena_interactive_2011}.
Other works leverage reduced dimension space to derive realistic animation parameters~\cite{lau_face_2009, cao_unsupervised_2003}.
Then, the final animation is reconstructed using linear interpolation or blending weights function.
The first works considering the temporal behavior of the face propose to propagate the edition by fitting a Catmull-Rom spline~\cite{li_orthogonal-blendshape-based_2008} or a B-spline curve~\cite{choe_performance-driven_2001} on the edited animation parameters.

Alternatively, more sophisticated interpolation methods were proposed such as a bilinear interpolation~\cite{arai_bilinear_1996}, spline function~\cite{kochanek_interpolating_1984} or cosine interpolation~\cite{parke_computer_1972}.
While easy to control and fast at generating coarse animation, the simplicity of the interpolation algorithms cannot mimic the complex dynamics of facial motions for segments longer than a few frames.
The resulting animation's quality is dictated by the number and relevance of user-created keyframes.

Seol and colleagues~\cite{seol_spacetime_2012} propagate edits using a movement matching equation.
In the same spirit, Dinev and colleagues~\cite{dinev_user-guided_2018} use a gradient-based algorithm to smoothly propagate sparse mouth shape corrections throughout an animation.
While producing high-quality results, their solutions rest on well-edited keyframes.
Ma et al.~\cite{ma_style_2009} learn the editing style on few frames through a constraint-based Gaussian Process and
then utilize it to edit similar frames in the sequence. Their methods are efficient at the time-consuming task of animation editing,
but it does not ensure temporal consistency of the motion.

To accelerate keyframe specification, several works explore methods to generate hand-drawn in-betweens~\cite{burtnyk_computer_1975} automatically.
Recently, considering that human motion dynamics can be learned from data, Zhang et al.~\cite{zhang_data-driven_2018} learn inbetween patterns with an auto-regressive two-layer recurrent network to automatically autocomplete a hopping lamp motion between two keyframes.
Their system offers the flexibility of keyframing and an intelligent autocompletion learned on data, but does not address the case of long completion segments.
Zhou et al.~\cite{zhou_generative_2020} address this with a learning-based method interpolating motion in long-term segments guided by sparse keyframes.
They rely on a fully convolutional autoencoder architecture and demonstrate good results on full-body motions.
As they point out, using convolutional models for temporal sequences has drawbacks, as it hard-codes the model's framerate, as well as the time-window on which temporal dependencies in the signal are considered by the model (the receptive field of the network).
Our experience indicates that recurrent networks seem to obtain better results in that case of facial data. One reason might be, the facial motions tend to exhibit less inertia and more discontinuities, which are better modeled by recurrent models' ability to learn to preserve or forget temporal behavior at different time scales.

\subsubsection{Data-based Motion Prediction}
\label{subsec:motion_pred_rw}

Our work focus on generating new motion through context-aware learning-based methods.
Predicting context-aware motion is a recent popular topic of research.
Since the seminal work of~\cite{fragkiadaki_recurrent_2015} on motion forecasting,
an increasing amount of work has addressed learning-based motion generation~\cite{ruiz_human_2019, wang_combining_2019,butepage_deep_2017} using previous frames~\cite{martinez_human_2017}.
Early learning-based works rely on deterministic recurrent networks to predict future frames~\cite{fragkiadaki_recurrent_2015, jain_structural-rnn:_2016, martinez_human_2017}.
Overall, recent works turn toward generative frameworks that have demonstrated state-of-the-art results in motion forecasting~\cite{wang_combining_2019,ruiz_human_2019, zhou_generative_2020}.
Ruiz and colleagues~\cite{ruiz_human_2019} propose a fully convolutional generative image inpainting framework, to predict and denoise body-motion sequences.
They suggest to occluding the last part of the animation, or discrete spatiotemporal features (either joints or frames), and regenerate a realistic completed animation.
In this work, we enable the user to edit both short and long motion parts, anywhere in the sequence, and guiding the generation of the new sequence in various ways.
Their methods are very relevant to this work, yet they do not consider semantic guidance to control the generated animation.

\subsection{Motion Editing}
\label{sec:facial_edit}
Multiple works leverage existing data to synthesize temporal motion matching user's constraints.
A first group of methods derives from data a subspace of realistic motion and performs trajectory optimization, ensuring natural motion generation.
Stoiber et al.~\cite{stoiber_automatic_2008} create a continuous subspace of realistic facial expressions using AAM, synthesizing coherent temporal facial animation.
Akhter et al.~\cite{akhter_bilinear_2012} learn a bilinear spatiotemporal model ensuring a realistic edited animation.
Another group of solutions is the use of motion graph~\cite{kovar_motion_2002, zhang_spacetime_2004}, which considers the temporality of an animation.
Zhang et al.~\cite{zhang_spacetime_2004} create a Face graph to interpolate frames realistically. 
Motion graph ensures a realistic facial animation, but it requires high memory cost to retain the whole graph.

The first one to propose a fully learning-based human motion editing system is the seminal work of Holden et al.~\cite{holden_deep_2016}.
They map high level control parameters to a learned body motion manifold presented earlier by the same authors~\cite{holden_learning_2015}.
Navigating this manifold of body motion allows to easily alter and control body animations, while preserving their plausibility.
Recently, several works emphasize the realistic aspect of generated motion through generative and adversarial techniques~\cite{wang_combining_2019, habibie_recurrent_2017}.
Habibie et al~\cite{habibie_recurrent_2017} leverage a variational autoencoder to sample new motion from a latent space.
Wang et al.~\cite{wang_combining_2019} stack a "refiner" neural network over the RNN-based generator, trained in an adversarial fashion.
While an intuitive and high-level parametrization steering a body motion have generated a consensus, there is no such standard abstraction to guide facial motion.
Later, Berson et colleagues~\cite{berson_robust_2019} use a learning-based method to perform temporal animation editing, providing meaningful temporal vertex distances.
However, this work needs explicit temporal constraints at each frame to edit, precluding a precise  keyframe-level control.
In this work, we propose a new point of view: a generative method from none, discrete or semantic inputs.\\

\subsection{Facial reenactment}
\label{sec:facial_reen}
Our work is also related to the problem of video facial reenactment. 
Facial reenactment consists of substituting facial performance in an existing video with ones from another source and recomposing a new realistic animation.
Video facial reenactment has been an attractive area of research~\cite{kim_neural_2019, kim_deep_2018, fried_text-based_2019, thies_face2face:_2016,
garrido_automatic_2014} for the last decades.
One instance of facial reenactment is Visual Dubbing, that consists of modifying the target video to be consistent with a given audio track~\cite{suwajanakorn_synthesizing_2017, bregler_video_1997,garrido_vdub_2015, chang_transferable_2005}.
Fried and colleagues~\cite{fried_text-based_2019} propose a new workflow to edit a video by modifying the associated transcript.
The system automatically regenerates the corresponding altered viseme sequence using a two-stage method: a coarse sequence is generated by searching similar visemes in the video and stitching them together. Then a high-quality photorealistic video is synthesized using a recurrent neural network.
This work follows the general trend and exploits recurrent GAN architecture~\cite{kim_neural_2019, song_everybodys_2020} to produce realistic facial animation matching semantic constraints.
However, our work does not aim to improve the photorealism of synthesized facial performance but instead, focuses on supplying a versatile and global facial animation editing framework.
Indeed, facial reenactment is devoted to a particular facial animation editing scenario, in which either a semantic or source animation is available, preventing flexible and creative editing applications.

\section{Method}
\begin{figure}[t!]
  \centering
  \includegraphics[width=.48\textwidth]{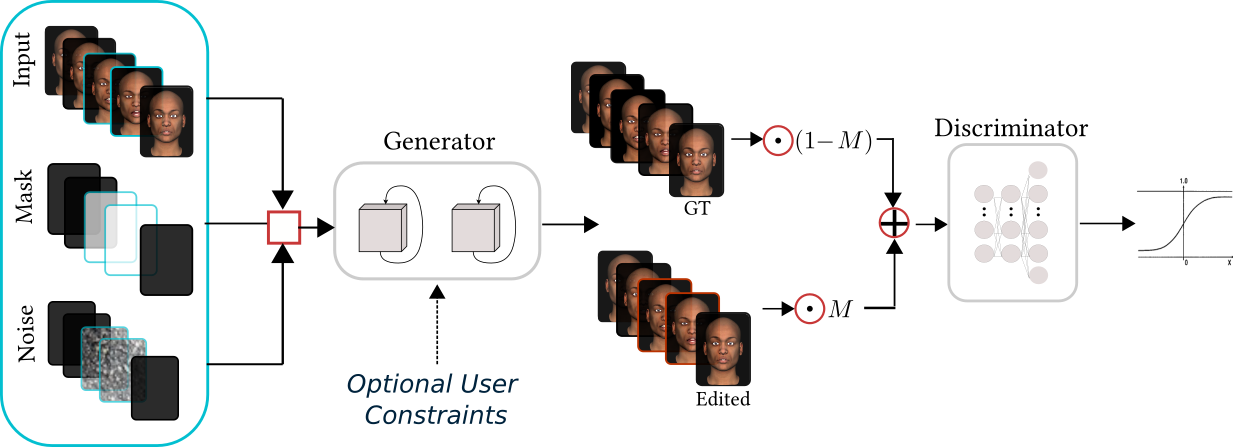}
  \caption{\
    \label{fig:overview}
Framework overview. We build our editing tool upon a GAN scheme, using an approach similar to image inpainting. We feed the generator with a mask, a masked animation and a noise vector, eventually we add constraints such as sparse keyframes, a noisy animation or sequence of visemes. The generator ends-up with the completed animation. The discriminator has to distinguish between real animation and fake ones: it is supplied with the ground-truth animation and the generated one (the partial ground-truth sequence completed with the generate parts).}
\end{figure}

Our goal herein is to train from data a generative neural network capable of generating plausible facial motions
given different kinds of input constraints such as sparse keyframes, discrete semantic input, or coarse animation.
In this section, we describe the parametrization of our system with the different constraints, enabling supervised motion editing (Section~\ref{sec:para_sys}).
We then detail our system based on the well-established GAN minmax game (Section~\ref{sec:archi_det}), as well as the training specifications.
An overview of our system is depicted in Figure~\ref{fig:overview}.

\subsection{Parametrization of our system}
\label{sec:para_sys}

Our system is meant to be used in any animation generation pipeline.
Therefore, we parametrize facial animations with the highly popular blendshape representation, common throughout academia and the industry~\cite{lewis_practice_2014}. 
We develop a framework similar to the image inpainting ones~\cite{yu_free-form_2019, jo_sc-fegan_2019}: more precisely, we consider an analogous training strategy for our networks. 
We feed a generator, $G$, with an incomplete animation, a noise vector, a mask
and optionally a discrete, noisy, or semantic input guiding the editing process.
At training time, the incomplete animation $\mathbf{X}_{i} \in \mathbb{R}^{L\times N}$ corresponds to the original ground-truth animation $\mathbf{X}_{gt} \in \mathbb{R}^{L\times N}$ with randomly erased segments signaled by the mask. Both the original and the incomplete input animations consist of the concatenation of $L=200$ frames of $N=34$ blendshape coefficients.
The mask $\mathbf{M} \in \mathbb{R}^{L\times N}$ encodes locations of erased segments (all blendshape coefficients) for a random number of consecutive frames. The input animation can be expressed as $\mathbf{X}_{i} = (\mathbf{1} - \mathbf{M})\odot \mathbf{X}_{gt}$.
$\mathbf{M}$ is a matrix with zeros everywhere and ones where blendshape coefficients are removed, and
$\mathbf{1}$ is an all-ones matrix of size $L\times N$.
The number and the length of masked segments in the input animation are chosen randomly, such as
at test time our network can edit both short and very long sequences.
At test time, masked segments are placed by the user to target the portions of the input sequence to edits.
We note that our network can also generate an animation by using a mask covering the full sequence.
The vector of noise, $\mathbf{z} \in \mathbb{R}^{L\times 1}$, is composed of independent components drawn from Gaussian distribution, with 0 mean and a standard deviation of 1.
We use the same framework for different editing scenarios and train a different network for each editing input type.
Our framework can also perform unguided motion completion in missing segments, which is useful in the case of long occlusions for instance.
In many cases though, the animator/user wants to guide the edit; so we focus on employing our framework for supervised motion editing.
To achieve this, we leverage the conditional GAN (CGAN)~\cite{mirza_conditional_2014} mechanism to add semantic guidance to our system.
We concatenate a constraint matrix to the input, $\mathbf{C}_{i} = \mathbf{\tilde{M}}_{i}\odot \mathbf{C}_{gt,i}$, with non-zero components where animation has been erased.
$\mathbf{C}_{gt,i} \in \mathbb{R}^{L\times N_{feat_{i}}}$ encodes the $i^{th}$ constraint vector of $N_{feat_{i}}$ features over time.
$\mathbf{\tilde{M}}_{i} \in \mathbb{R}^{L\times N_{feat_{i}}}$ is the constraint-specific mask matrix, with zeros everywhere and ones at the same frame indices as $\mathbf{M}$.
    The constraints $\mathbf{C}_{gt,i}$ can be a sparse matrix of keyframes, a dense noisy animation, or one-hot vectors representing pronounced visemes at each frame.
Each constraint conditions the training of the corresponding specific system.
We consider three high-level constraint types enabling animation editing for several use cases:
\begin{itemize}
    \item \textit{Keyframes}:
        One main cause of animation editing is expression modifications, such as correcting the shape of the mouth or adding new expressions. Hence, we add sparse keyframes extracted from the ground-truth animation as constraints. The time between two keyframes is chosen randomly between 0 and 0.8 seconds.
\item \textit{Noisy animation}:
    Our system enables the user to change the content of the animation and guide it with a coarse animation, such as one obtains from consumer-grade motion capture on consumer devices (webcam, mobile phone, ...).
\item \textit{Visemes}:
    We also consider a more semantic editing use case, such as speech corrections from audio.
        We use an audio-to-phoneme tool to obtain annotation in phonemes of each sequence in the database.
        In this work, we use the Montreal-Forced-Aligner~\cite{mcauliffe_montreal_2017}, but any audio-to-phoneme tool can be used.
        We constrain our network with a one-hot vector representing the visemes at each time.
        A viseme is the visual facial representation of a group of phonemes.
        We group all phonemes in 18 classes of visemes presented in Table~\ref{tab:map_phnm}.
   
\end{itemize}

\begin{table}
    \caption{\label{tab:map_phnm}
    Groups of phonemes.}
    \begin{tabular}{l|l||l|l}
        Visemes            & Phonemes                             & Visemes                         & Phonemes\\
        \hline
        sil               &                                      & G + K + H                      & g, k, q, \textscg \\
        AO + OY           & a, \textopeno                        & L + N + T + D                  & l, n, t, d, \textscl, \textdoublebaresh, \textfishhookr \\
        AA + AE + AY      & \ae, \textscripta                    & S + Z                          & s, z, \textgamma  \\
        EH + EY           & e, \textepsilon, e\textsci           & Sh + Ch + Zh                   & \textesh, \textteshlig, \textyogh \\
        IH + IY + EE + IX & i, \textsci, \textbari               & TH + DH                        & \texttheta, \dh \\
        OH + OW           & o, \textturnscripta                  & F + V                           & f, v \\
        AH + ER           & \textturnv, \textschwa, \textrhookschwa, \textrhookrevepsilon & M + B + P & b, m, p \\
        UW + AW + UH      & u, \textupsilon, a\textupsilon       & W                              & w, \textturnw \\ 
        JH                & j, \textdyoghlig                     & R                              & \textturnr\\

\end{tabular}
\end{table}

\subsection{Framework details}
\label{sec:archi_det}
We consider a generative approach relying on the well-known GAN principle.
Hence, as in any GAN framework, our system is composed of two neural networks:
a generator, designed to fill the timeline with realistic animation, and a discriminator intended to
evaluate the quality of the generated animation.

Our generator, $G$, has to learn the temporal dynamics of facial motion.
We use a recurrent architecture for our generator, as sharing parameters through time have demonstrated impressive results in modeling, correcting, and generating intricate temporal patterns.
Our generator uses a Bidirectional Long Short-Term Memory (B-LSTM) architecture for its capability to adapt to quickly changing contexts yet also model long-term dependencies.
Our generator consists of a sequence of $N_{layers}$ B-LSTM layers ($N_{layers}$=2) with a stacked final dense output layer to get dimensions matching the output features.
The recurrent layers consist of 128 hidden units.
The main goal of the generator is to create plausible animations, i.e., to fill a given timeline segment with realistic motion signals that smoothly connects to the motion at the edge of the segment.

Our discriminator, $D$, has to learn to distinguish between a generated animation and a one produced by ground-truth motion capture.
Because we want our generator to create an animation that blends well outside its segment, we supply our discriminator with the entire animation rather than only the generated segment, and choose a convolutional structure for $D$.
Some elements have a higher impact on the quality perception of a facial animation.
For instance, inaccuracies in mouth and eye closures during speech or blinks are naturally picked up as disturbing and unrealistic.
Thus, we enrich the discriminator's score with relevant distance measurements over time that matches those salient elements.
Our discriminator's structure is inspired by recent advances in image inpainting~\cite{yu_free-form_2019, jo_sc-fegan_2019}.
It is a sequence of 4 convolutional layers, followed by spectral normalization~\cite{miyato_spectral_2018},
stabilizing the training of GANs. Over the convolutional layers, we stack a fully connected layer predicting the plausibility of the input animation. The convolutional layers get a kernel of size 3, scanning their input with a stride of 2, and end up with respectively 64, 32, 16, 8 channels. We use the LeakyRelu activation function~\cite{xu_empirical_2015} after every layer except the last one.

\subsection{Training methods}
\label{sec:train_met}

Classically, to train the proposed system we consider the minmax game between the generative and the discriminative losses.
The generative loss is inspired by~\cite{jo_sc-fegan_2019, yu_free-form_2019}, and it is the sum of three terms.
Our generator has to reproduce the input animation outside masked segments faithfully.
Thus, we define a loss ensuring accurate animation reconstruction:
\begin{equation}
    \mathcal{L}_{feat} = \alpha_{gt}(\mathbf{1}-\mathbf{M})\odot|G(\mathbf{X}_{i}) - \mathbf{X}_{gt}| + \mathbf{M}\odot|G(\mathbf{X}_{i}) - \mathbf{X}_{gt}|
\end{equation}
The blendshape representation weights salient shapes such as shapes controlling eyelid closure and shapes with minor effect such as the one affecting the nose deformation equally.
As Berson et al.~\cite{berson_robust_2019}, we add a loss $\mathcal{L}_{dis}$, to focus preservation of some key inter-vertices distances between the estimate and the ground truth animations. 
$\mathcal{L}_{dis}$ encourages accurate mouth shape and eyelid closure, crucial ingredients for realistic facial animation. It focuses on six distances: the first three measure the extent between the upper and lower lips (at three different locations along the mouth), the fourth is the extent between the mouth corners and the last two measure the opening of the right and left eyelids.
Finally, the generator is trained to minimize the following loss:
\begin{equation}
    \mathcal{L}_{G} =  \mathbb{E}[1 - D(G(\mathbf{X}_{i}))] + w_{feat}\mathcal{L}_{feat} + w_{dis}\mathcal{L}_{dis}.  
\end{equation}

At the same time, we train our discriminator to minimize the hinge loss.
We force the discriminator to focus on the edited part by feeding it with 
a recomposed animation $\mathbf{X}_{rec}$, which is the incomplete input animation completed with the generated animation, i.e, $\mathbf{X}_{rec} = (\mathbf{1}-\mathbf{M})\odot \mathbf{X}_{gt} + \mathbf{M}\odot G( \mathbf{X}_{i})$.
We also influence the discriminator attention by providing it the key intervertices distances mentioned earlier.
We add the WGAN-GP loss~\cite{gulrajani_improved_2017}, $\mathcal{L}_{gp} = \mathbb{E}[||(\nabla_{\mathbf{U}}D(\mathbf{U})\odot \mathbf{M}||-1)^2]$ to make the GAN training more stable.
In this formula, $\mathbf{U}$ is a vector uniformly sampled along the line between discriminator inputs from $\mathbf{Y}_{gt}$ and $\mathbf{Y}_{rec}$, i.e, $\mathbf{U} = t\mathbf{Y}_{gt} + (1-t)\mathbf{Y}_{rec}$ with $0 \leq t \leq  1$.
Hence, the loss of the discriminator is:
\begin{equation}
    \mathcal{L}_{D} =  \mathbb{E}[1 - D(\mathbf{Y}_{gt})] + \mathbb{E}[1 + D(\mathbf{Y}_{rec})] + w_{gp}\mathcal{L}_{gp},
\end{equation}

where $\mathbf{Y}$ refers to the concatenation of an animation and its corresponding intervertices distances.
For all our experiments, we set $w_{feat}= 1$, $\alpha_{gt} =10$, $w_{gp} = 10$ and $w_{dis}=1$.
We set the initial learning rate of both the generator and the discriminator at 0.001.
We use the Adam optimizer~\cite{kingma_adam:_2014}.
We add a dropout of 0.3 to regulate the generator.
This system has been implemented using the Pytorch framework.

\section{Results}
In this section, we demonstrate our system's capability to render realistic animation with different types of editing constraints.
First, we detail the data used for the training and the testing of our framework (Section~\ref{sec:gathered_data}).
Then, we describe the different scenarios in which our framework might be useful, from unsupervised motion completion (Section~\ref{sec:unsupervised}),
to constraint-based motion editing (Section~\ref{sec:supervised}).

\subsection{Gathered Data}
\label{sec:gathered_data}
We use two datasets for our experiments.
We leverage the enhanced audiovisual datasets "3D Audio-Visual Corpus of Affective Communication" (B3D(AC)\^{}2)~\cite{fanelli_3-d_2010, berson_robust_2019} to train our networks, especially the one requiring both facial animation and phoneme labels.
Overall, the corpus amounts to 85 minutes of animation and will be released for reproducibility of our results.
We add another dataset, which consists of performance-based animations, manually created with a professional animation software.
From the original videos, we also employ an automatic face tracking solution to generate coarse, noisy animation corresponding to those videos.
Those trackers are noisy by nature, so we do not need to add artificial noise to the input.
We use this last dataset alone to train our "noisy-signal-based" editing system.
This training set contains 52 sequences (49 minutes of animation) recorded at different framerates between 30 and 120 frames-per-second (fps).
For all our experiments, we resample every animation at 25 fps (the framerate of the (B3D(AC)\^{}2) dataset) and use the same blendshape model counting 34 blendshapes for every animation of each of our scenarios.

As with any learning-based methods, it is essential to know how the proposed approach depends on the training data.
To test our framework, we record new sequences with a different subject, reciting new sentences, and performing different expressions to check if the model generalizes well.
We derive both the original animations and the noisy ones using the same procedure as described above.

\subsection{Unsupervised Motion Filling}
\label{sec:unsupervised}
First, we demonstrate the capability of our system to generate plausible animation without any supervision.
We validate our system using animation of the test set by randomly removing some parts of them.
We regenerate a complete sequence using our network, producing undirected motion filling.
As we can see in the accompanying video, the generated parts (lasting 2.6s) are blended realistically with the animation preceding and following the edit.
In this sequence, our generator produces "talking-style" motions and hallucinates eyebrows movements rendering the edited parts more plausible.

\begin{figure}[t]
  \centering
  \includegraphics[width=.45\textwidth]{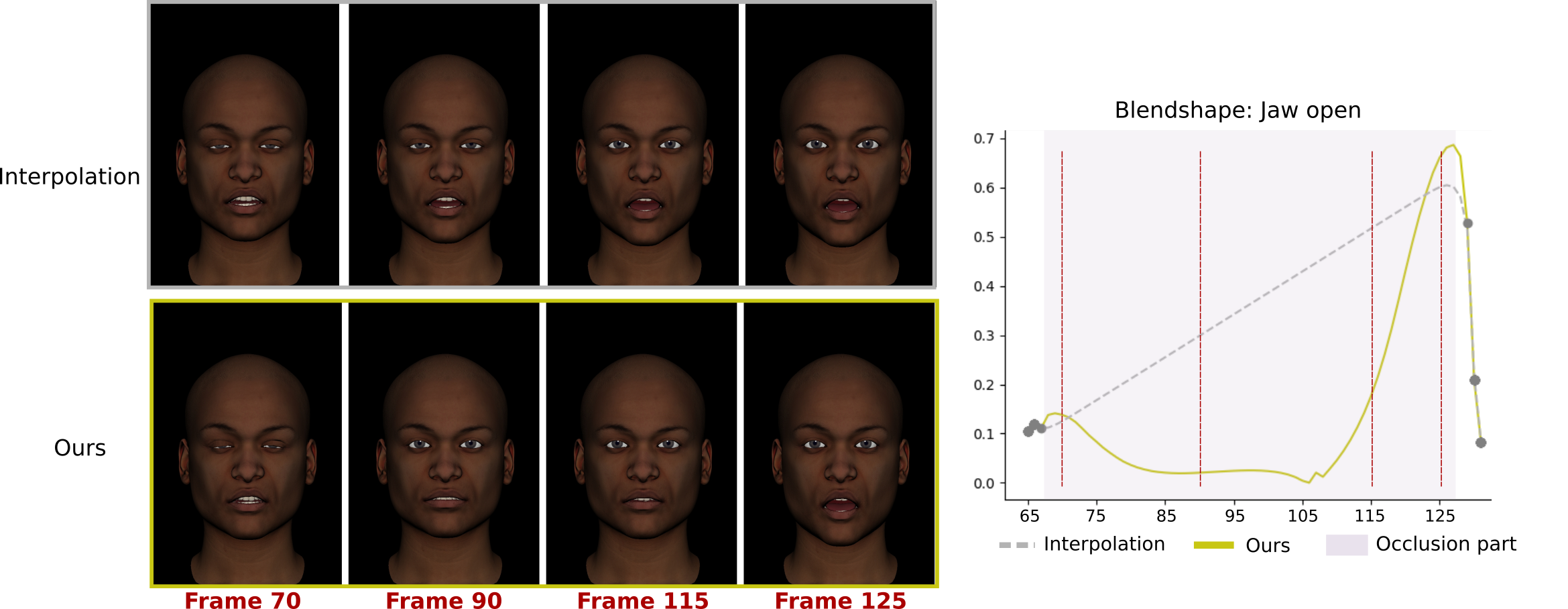}
  \caption{\
    \label{fig:occlusion_comp_lin}
Occlusion motion completion. Compared to standard linear interpolation solving, our system generates realistic motion dynamics: in case of long occlusions, our system ensures that the mouth returns to the neutral poses. Moreover, as we use a bidirectional architecture, our system anticipates the wide opening of the mouth and smoothly re-open the mouth from the neutral pose.}
\end{figure}

One potential application of our unsupervised animation generation system is its capability to generate
more realistic sequences in case of long occlusions than simple interpolation methods.
We use a new recorded sequence with occlusion of around 3 seconds (about 75 frames).
Such occlusions often alter the quality of the final animation and require manual cleaning.
We compare our generative system with a sequence resulting in interpolating the missing animation with boundaries and
derivatives constraints.
As we can see in Figure~\ref{fig:occlusion_comp_lin}, filling the gap with interpolation
leads to long oversmoothed motions, far from realistic motion patterns.
Our system creates a more realistic sequence: the subject first returns to the neutral pose and anticipates the wide mouth opening by smoothly reopening the mouth.
One might also observe the eyebrows activation, consistent with the mouth openings.

\subsection{Guided Motion Editing}
\label{sec:supervised}

While unsupervised motion completion can be used to handle long occlusions, most relevant uses require users to steer the editing process.
In the following, we present several use cases of guided facial animation editing.
We test our system using both the test set, which is composed of sequences of unseen subjects, and
new performance-based animations recorded outside the dataset.

\subsubsection{Keyframes}

\begin{figure}
      \centering
  \includegraphics[width=.45\textwidth]{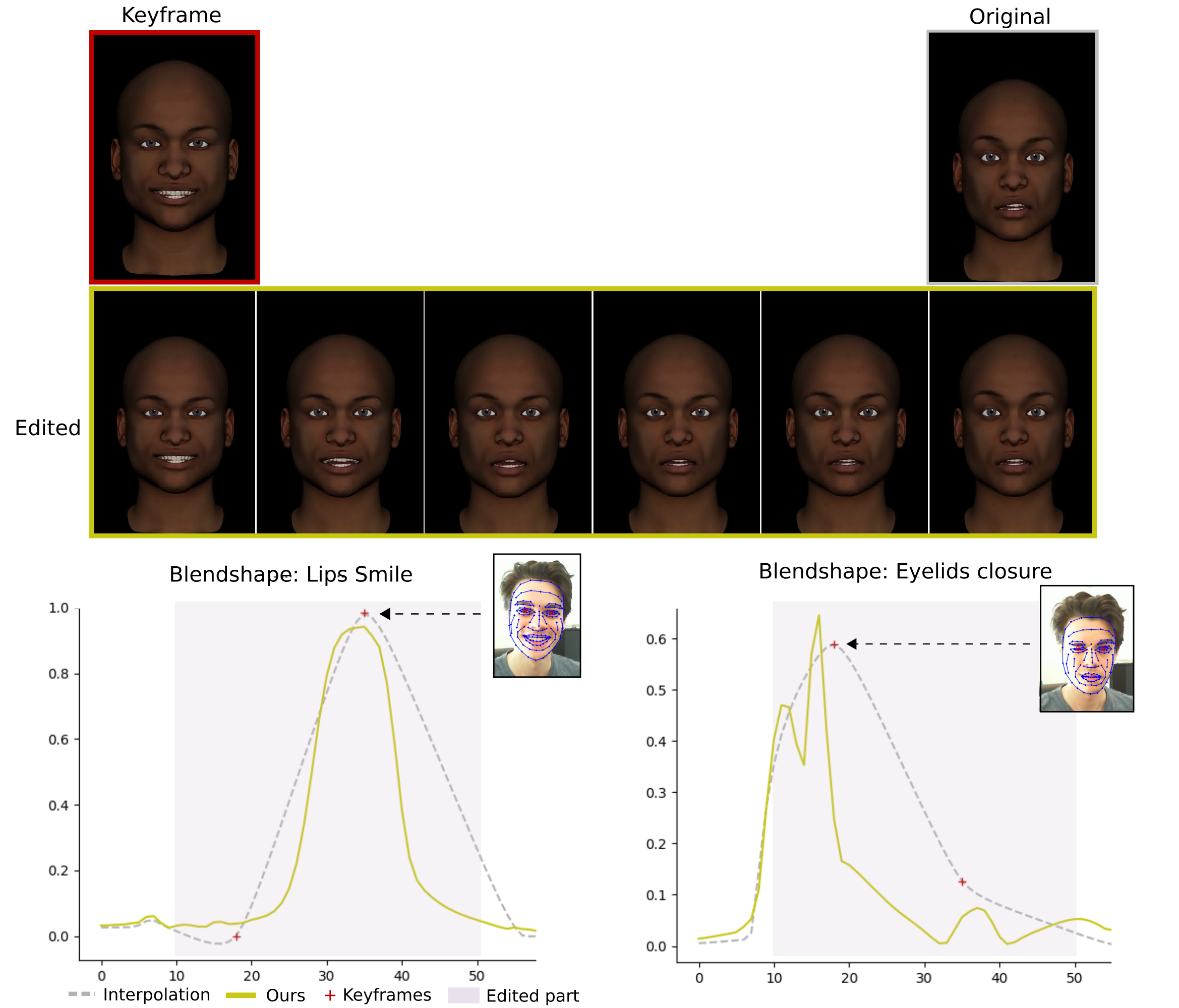}
  \caption{\
      \label{fig:key_f_marc}
Validation of our keyframes-based constraints system on our testset with new coarse animation. Our system ensures natural coarticulation between key frames constraints and input signal.}
\end{figure}

It is common for performance-based animation to require additional or localized corrections either due to technical or artistic considerations.
Ideally, one would simply use new captured or hand-specified expressions to edit the animation and expect the editing tool to derive the right facial dynamics, reconstructing a realistic animation automatically.
This use case has motivated the keyframe-based supervision of our editing system.
We test our system's ability to handle this scenario by randomly removing parts of the input animation and inputting the network with sparse, closely- or widely-spaced, keyframe expressions.
We observe that the system outputs natural and well-coarticulated motions between the keyframe constraints and the input signal:
as we can see in Figure~\ref{fig:key_f_marc}, our system generates non-linear blending around the smile keyframe expressions, and naturally reopens the mouth at the end of the edited segment.
We can see in the video that our system generates a more natural and organic facial dynamics than classic interpolation.

\begin{figure}[t]
  \centering
    \begin{subfigure}{.45\textwidth}
      \centering
  \includegraphics[width=.98\textwidth]{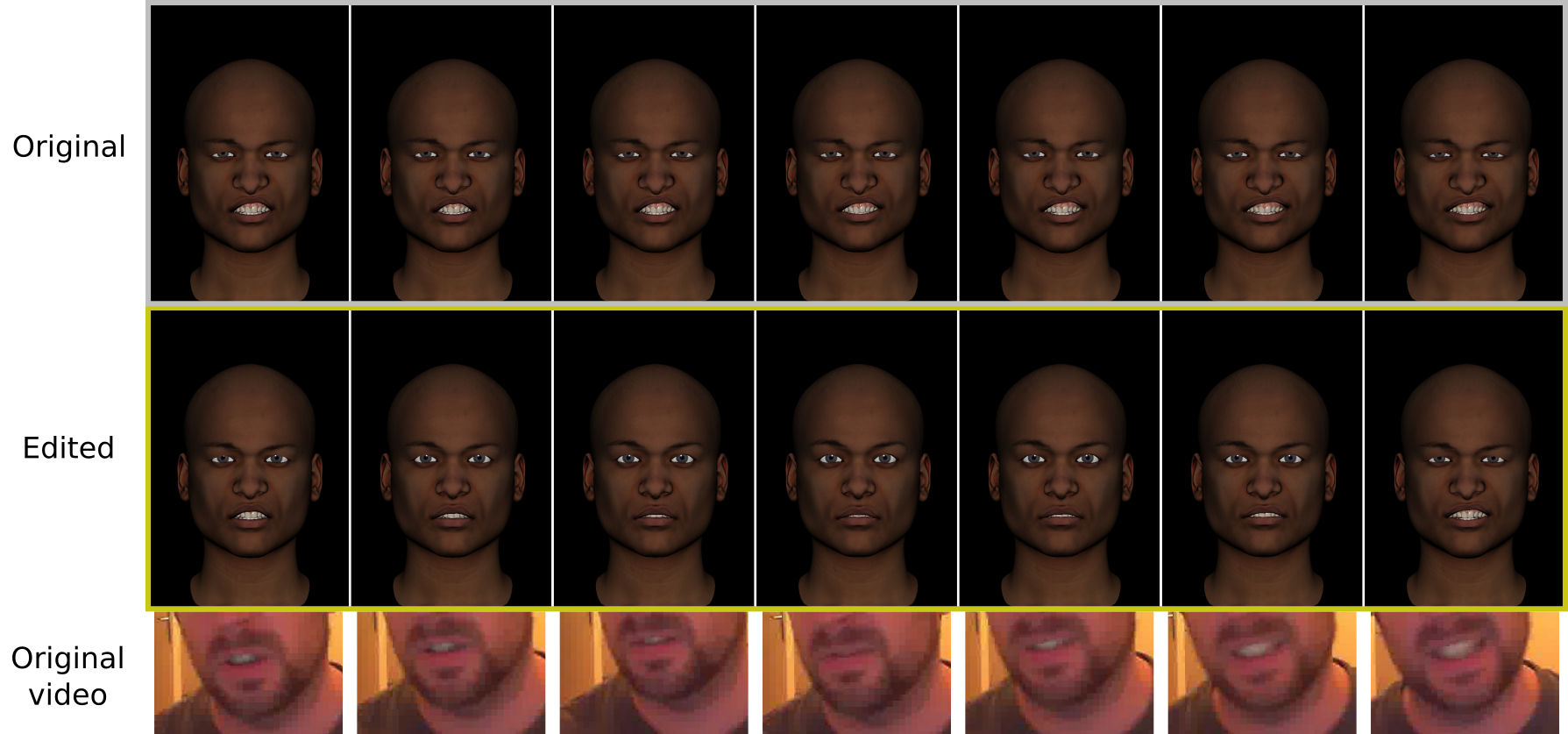}
  \caption{\
    \label{fig:key_close_mouth}
    Modification of the mouth shape. Our system generates a more faithful shape of the mouth, given only one keyframe.}
        \end{subfigure}
        \newline
   \begin{subfigure}{.45\textwidth}
      \centering
  \includegraphics[width=.98\textwidth]{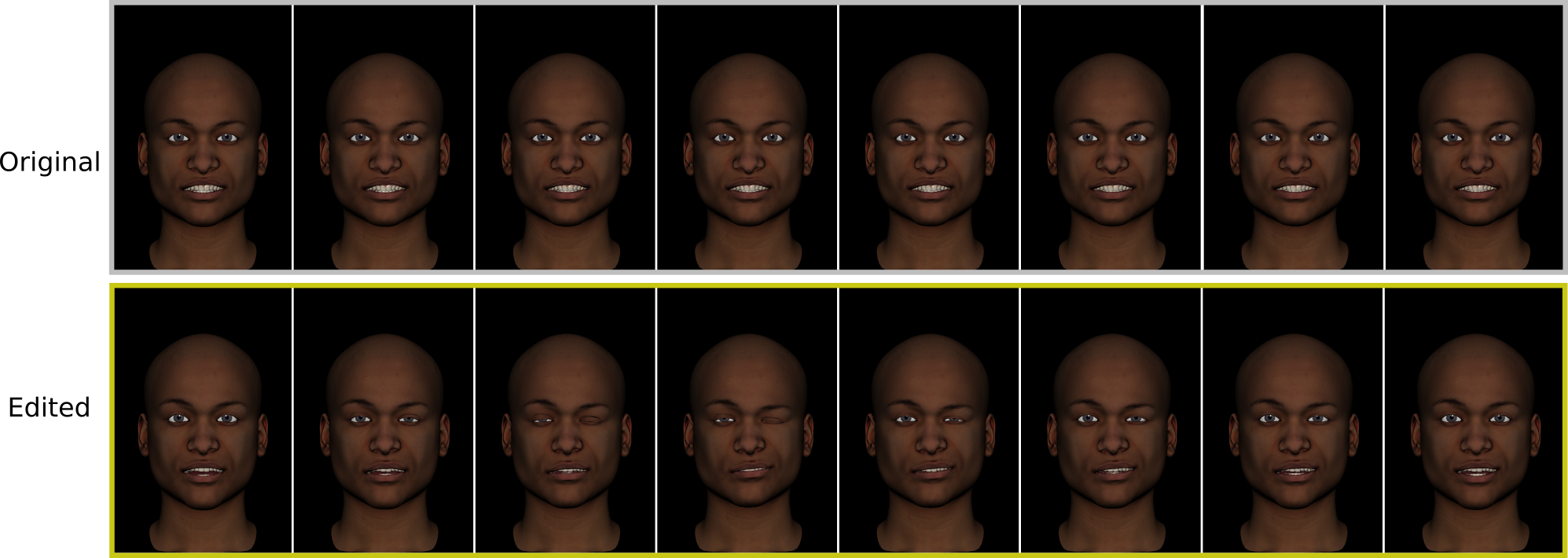}
  \caption{\
      \label{fig:key_f_wink}
Addition of one expression such as a wink. Our system naturally adds a key-expression: as we can observe, the mouth motion consistently moves to re-match the smiling expression.}
        \end{subfigure}
  \caption{\ Keyframe-based Editing. Our system generates realistic motions with only a few keyframes as a constraint.
  \label{fig:key_f_edition}}
\end{figure}

Another use case is adding an expression not present in the existing animation. 
For instance, in one of our videos, the performer forgot the final wink move at the end of the sequence (see~\ref{fig:key_f_wink}).
We simply add it to the sequence by constraining the end of the sequence with a wink keyframe, which has been recorded later.
We can observe in Figure~\ref{fig:key_f_wink} how naturally the mouth moves to combine the pre-existing smiling expression and the added wink request. 

Finally, one recurrent shortcoming of performance-based animation is getting a mouth shape that does not match the audio speech.
For instance, on a video outside the dataset, we observe that the face capture yields imprecise animation frames of the mouth.
As we can see in Figure~\ref{fig:key_close_mouth}, the mouth should be almost closed, yet it remains wide open during a few frames.
We fed the desired expression as a keyframe input to the system, and let the system generate the corrected mouth motion~\ref{fig:key_close_mouth}.
The visual signature of labial consonants is a mouth closure. 
In the same editing spirit, our system can revise an inaccurate labial viseme by imposing mouth closure.
We display an example of this correction in the accompanying video.

\subsubsection{Noisy Animation}
\begin{figure}
      \centering
  \includegraphics[width=.45\textwidth]{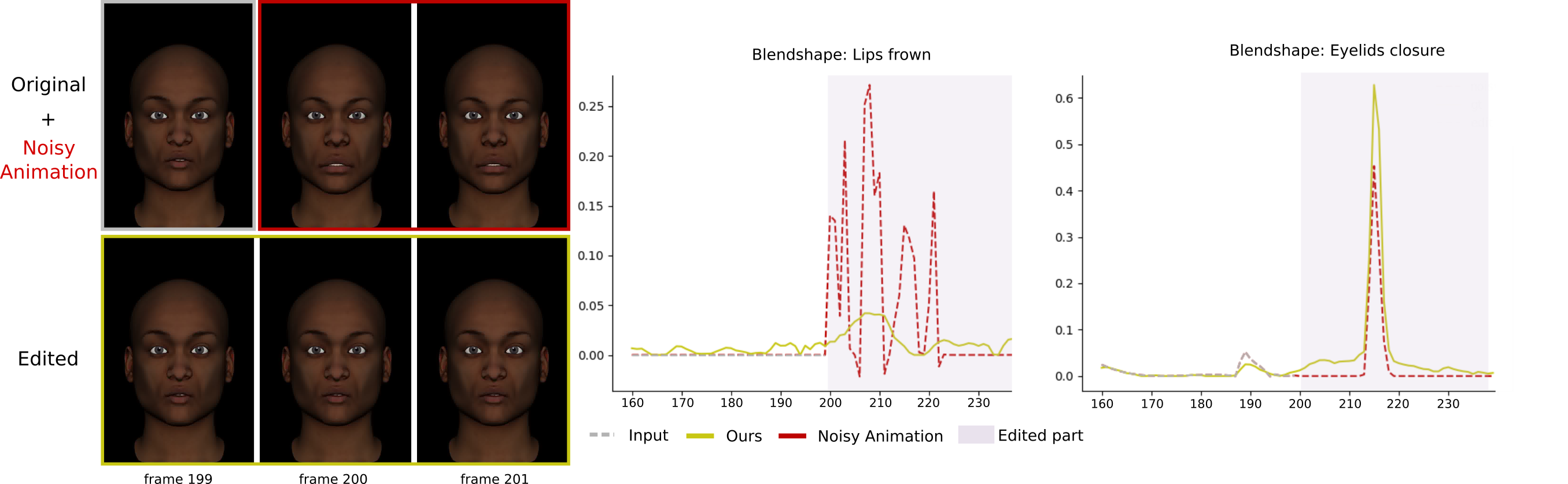}
  \caption{\
      \label{fig:noisy_input}
  Noisy animation-based system. We mask half the original sequence and feed the network with the other half noisy animation. As we can see on the left, our system removes jitters and unnatural temporal patterns, generating a smooth animation at the boundary. We can see on the right, how the unrealistic lips frowning movements are filtered by our system, while the natural dynamic of the eyelids is preserved.}
\end{figure}

Animation changes longer than a few seconds would require specifying many guiding keyframes.
Instead, when long segments need to be substantially changed one could guide animation editing with lower-quality facial tracking applications, using webcam or mobile phone feeds.
In that case, the guiding animation is noisy and inaccurate, but is a simple and intuitive way to convey the animation intent.
We test this configuration, feeding our system with noisy animations generated from a blendshape-based face tracking software as a guide for the animation segment to edit.
As we can see in Figure~\ref{fig:noisy_input}, our system removes jitters and unrealistic temporal patterns but preserves natural high-frequency components such as the eyelids closures.

\subsubsection{Visemes}

\begin{figure}[t]
  \centering
  \includegraphics[width=.45\textwidth]{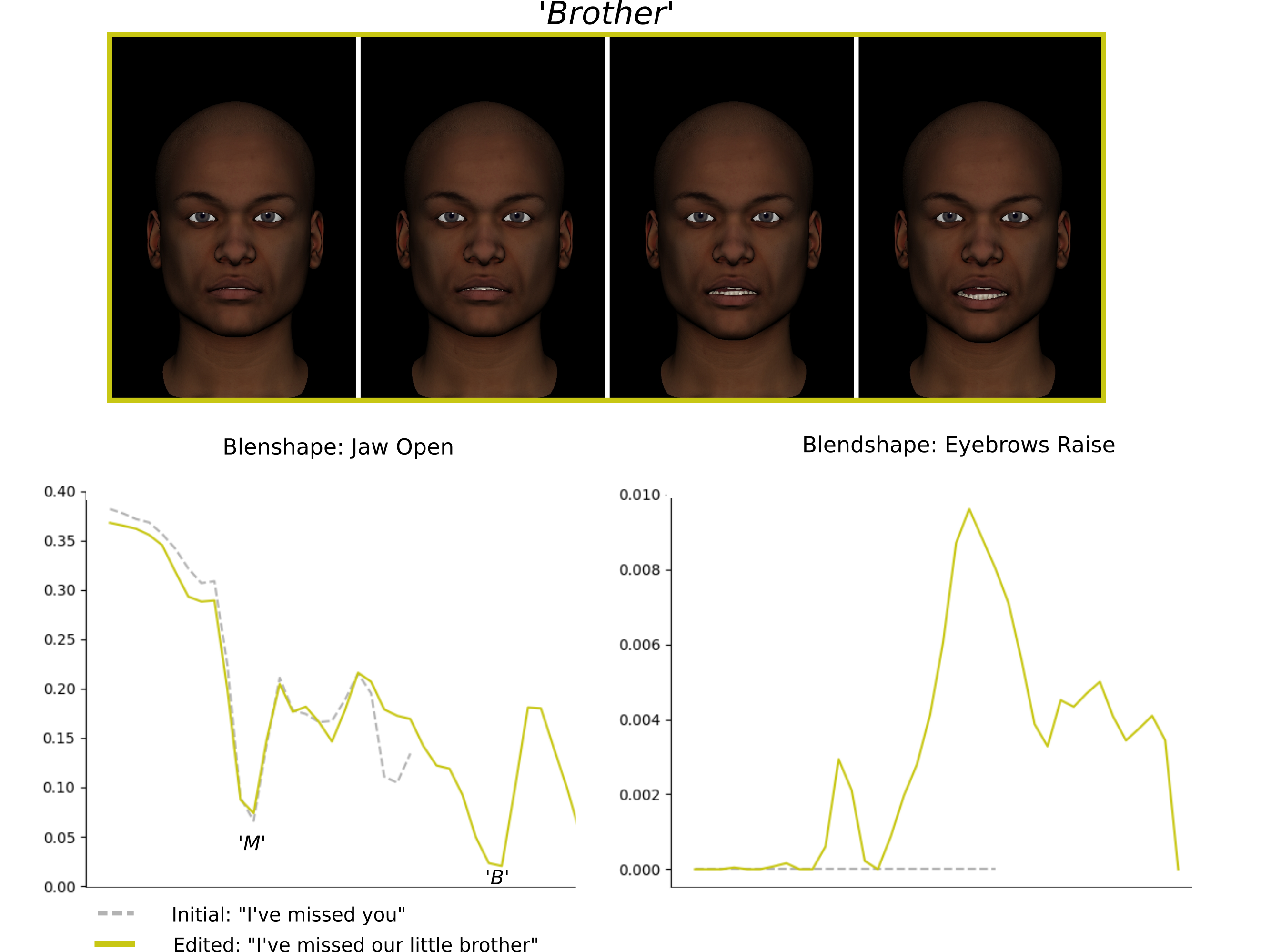}
  \caption{\
      \label{fig:brother_pro}
      Our system modifies the jaw motion according to the input constraints such as adjusting the jaw opening to fit bilabial consonant constraints. It hallucinates micro-motions such as raising eyebrows to make the editing part more plausible. (Left) Generated frames given the input phonemes sequence \textit{"brother"}.}
\end{figure}

We demonstrate the capability of our system to edit an animation semantically.
We use the initial sentence found in the test set
\textit{"Oh, I've missed you. I've been going completely doolally up here."}.
We generate a new animation by substituting 
\textit{"you"} with other nouns or noun phrases pronounced by the same subject in order to have consistent audio along with the animation.
As we can see in Figure~\ref{fig:brother_pro}, our system generates new motions consistent with the input constraints, "our little brother": it adjusts the movements of the jaw to create a realistic bilabial viseme. We observe the closure of the mouth when pronouncing \textit{"brother"} in Figure~\ref{fig:brother_pro}.
It hallucinates consistent micro-motions, such as raising eyebrows at the same time, favoring natural-looking facial animation.
Other examples are shown in the supplementary video.

\begin{figure}[t]
  \centering
  \includegraphics[width=.45\textwidth]{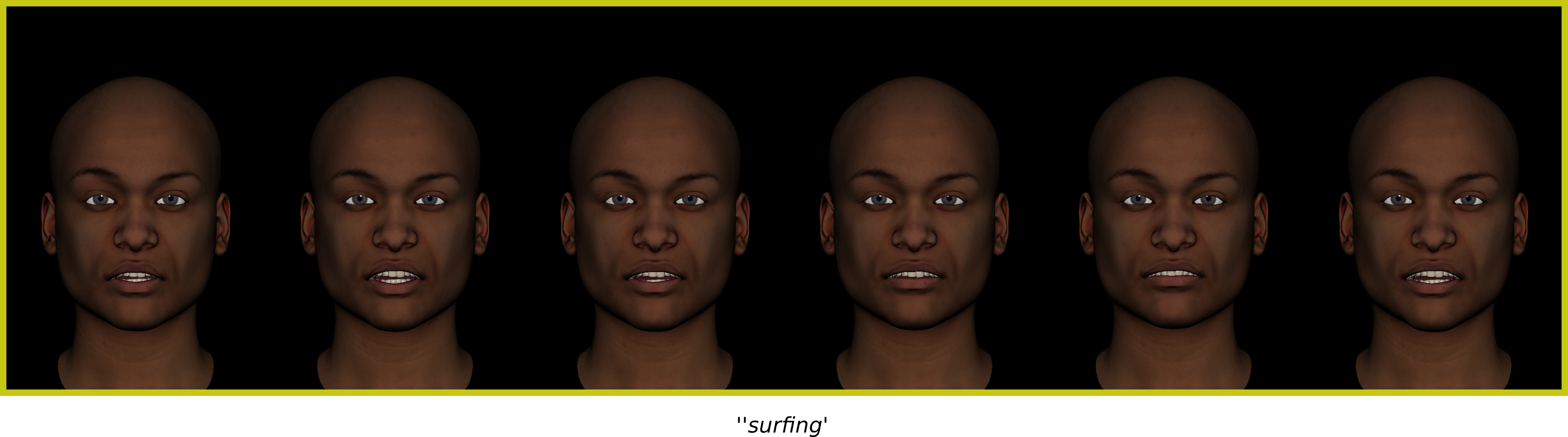}
  \caption{\label{fig:surfing}
      Generated frames given the input phonemes sequence \textit{"surfing"}.
      We can notice the bottom lip raises up to the bottom of the top teeth to generate the viseme \textit{"f"}.}
\end{figure}

We also perform viseme-based editing on a new subject reciting new sentences. 
For instance, we turn the initial sentence \textit{"My favorite sport is skiing. I’m vacationing in Hawai this winter.”}
into \textit{"My favorite sport is surfing. I’m vacationing in Hawai this winter.”}
The generated motion follows the new visemes sequence \textit{"surfing"} in Figure~\ref{fig:surfing}.
More precisely, we can see the bottom lip raising up to the bottom of the top teeth to generate the viseme \textit{"f"}.

\section{Evaluation}
In this section, we present quantitative evaluations of our framework.
First, we demonstrate the capability of our approach to reduce the manual effort required to edit facial animation. 
We then compare our methods with related ones in dealing with controllable animation editing.
Finally, we assess the quality of our results by gathering user evaluation on a batch of edited sequences.

\subsection{Fast Animation Editing System}
The principal objective of this work is to provide a system that accelerates the editing task.
We timed two professional animators to measure the average time they need to create a sequence of 100 frames (see Table~\ref{tab:time_artists}).
From this experiment, we find that it takes between 20 and 50 minutes to create a 100-frames animation, depending on the complexity and framerate of the animation. 
This amounts to an average individual keyframe setup time between 12 and 30 seconds.
We note that this estimation is consistent with the study conducted by Seol et al.~\cite{seol_artist_2011}.
Now, for different sequences processed by our method, we estimate the number of keyframes that would be required to produce the same result manually.
We proceed as follows: we automatically estimate the number of control points necessary for a cubic B\'{e}zier Curve fitting algorithm to approximate
the edited animation curves, within a tolerance threshold set at 0.01.
This process is repeated for each animation parameter independently.
In manual facial animation, some complex motions require very dense keyframe layouts to look realistic, making our method all the more appealing.

\begin{table}
    \centering
    \caption{\label{tab:time_artists} Average time to create 100-frames animation.}
    \begin{tabular}{l|c|c}
             & Handmade (min) & With animation software (s) \\
             \hline
        Artist 1 & $\sim$ 20 & 20  \\
        Artist 2 & $\sim$ 50 & 60 \\
\end{tabular}
\end{table}

\begin{table*}
    \centering
    \caption{\label{tab:time_estimation} Time performance evaluation. We compare the time to edit few animation with our system and manual keyframing. Our system considerably reduces the time of facial animation editing.}
     \begin{tabular}{l|c|c|c|c|c|c}
                             & \# of      & \# of estimated& Average error & Manually created & Inference full \\
                             & frames     & B\'{e}zier points  & by parameters & by an animator   & sequence (CPU) \\
        \hline
        Occlusion completion & 62         & 36            & 0.01          & $\sim$ 12 min    &0.14s\\
        Viseme editing       & 19         & 15            & 0.012         & $\sim$ 5 min    & 0.12 s\\
        Noisy-based          & 116        & 93            & 0.01          & $\sim$ 31 min    & 0.12 s \\
    
    \end{tabular}
\end{table*}

We compare in Table~\ref{tab:time_estimation} the time to edit a few animations with our system and manual keyframing.
From this experiment, we note that our system considerably reduces the time required to edit animation segments.

\subsection{Comparison with Continuous Control Parameters Editing Systems}
Recent controllable motion generation studies have an objective akin to animation editing, as they use regression neural networks to generate motion from high-level inputs.
We compare our system to two previous works, closely related to motion editing: the seminal work of Holden et al.~\cite{holden_deep_2016} on controlled body motion generation, and the recent work on facial animation editing of Berson et al.~\cite{berson_robust_2019}.
For a fair comparison, we use the same control parameters as~\cite{berson_robust_2019}, and regress the corresponding blendshape weights using either the fully convolutional regressor and decoder of~\cite{holden_deep_2016}, or the 2-network system proposed by~\cite{berson_robust_2019}.
We quantitatively compare the reconstruction error between these methods and our system on the test set. 
Therefore, we mask-out the complete input animation and feed our network with the control parameter signals.
We measure the mean square error between the original animation and the output one.
As we can see in Table~\ref{table:mse_comp}, our system achieves better performances than a regression network trained with MSE only.  
\begin{table}[H]
    \centering
    \caption{\label{table:mse_comp} MSE between high level parameters and our network with 8 control parameters.}
    \begin{tabular}{l|c|c}
                                & $MSE$   \\
             \hline
     \cite{holden_deep_2016}    & 0.016       \\
     \cite{berson_robust_2019}  & 0.018         \\
     Ours                       & \textbf{0.014}          \\
\end{tabular}
\end{table}

\begin{figure*}[t!]
  \centering
  \centering
    \begin{subfigure}{.95\textwidth}
      \centering
  \includegraphics[width=.95\textwidth]{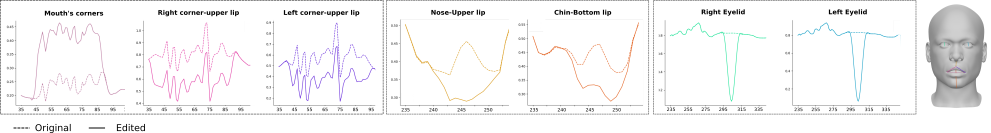}
  \caption{\
    \label{fig:cp_edit_curve}
We manually deform control parameter's curves 3 times (Top). The corresponding control parameters are displayed on the right-hand side.}
        \end{subfigure}
        \newline
   \begin{subfigure}{.95\textwidth}
      \centering
  \includegraphics[width=.95\textwidth]{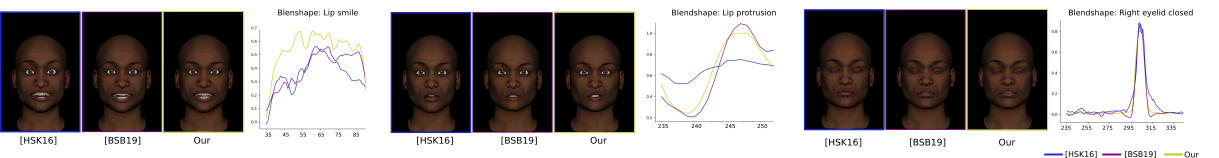}
  \caption{\
  \label{fig:comp_cp_def}As we can observe, our system generates realistic expressions consistent with the input constraints such as the regression-based systems developed by Holden et al.~\cite{holden_deep_2016} and Berson et al.~\cite{berson_robust_2019}. Stretching control parameter curves to match sparse constraints may yield unrealistic control parameter trajectories. However, our generative approach always generates motion with realistic dynamics.}
        \end{subfigure}
        \caption{\ Comparison with controllable motion editing systems. \label{fig:comp_prev}}
    \end{figure*}

We also observe qualitative differences between regressors~\cite{holden_deep_2016,berson_robust_2019} and our current approach. We do so by feeding our generator with dense control parameter curves, as used by regressors (see Figure~\ref{fig:cp_edit_curve}).
Even when stretching and deforming control curves to match sparse constraints, our system robustly continues to generate animation with realistic dynamics (Figure~\ref{fig:comp_cp_def}).
As mentioned by Holden et al.~\cite{holden_deep_2016}, the main issue with regression frameworks is the ambiguities of high-levels parameter inputs: the same set of high-level parameters can correspond to many different valid motion configurations.
We test the behavior of our approach in such ambiguous cases, by using very few input control parameters (3): the mouth opening amplitude, the mouth's corners distance, and one eyelids closure distance.
We indeed observe that a more ambiguous input signal leads to a noisier output animation for regression networks.
With the same input, our system is able to hallucinate missing motion cues outputs, producing a more natural and smooth animation.
We note that our system is even capable of creating plausible dynamics for the whole face in an unsupervised fashion (Section~\ref{sec:unsupervised}).

\subsection{User Feedback}

One widely recognized issue with animation generation methods is reliable evaluation of animation quality. 
Indeed, there is no quantitative metrics that reflect the naturalness and the realism of facial motions.
Hence, we gather qualitative feedback on edited animation generated by our system in an informal study.
A sample of 44 animation sequences -with different lengths and with or without audio- were presented to 21 subjects. 
Half the animations were edited with our system, using either visemes constraints, keyframes expressions, noisy signals, or in an unsupervised fashion.
Subjects were asked to assess whether the animation cames from original mocap or was edited.
In essence, participants were asked to play the role of the discriminator in distinguishing original from edited sequences. 
Most of the participants were not accustomed to close observation of 3D animation content. 
We gather the following user feedback among the 21 subjects: 54\% of the original animations were classified as such (true positive), while 51\% of edited sequences were also classified as original ones (false positive).
We also show the sequences to 5 experienced subjects, that know the context of this work: even they detected only 58\% of the edited sequences (true negative) and half of the original ones (true positive).

\section{Conclusion and Future Work}

We have proposed a generative facial animation framework able to handle a wide range of animation editing scenarios. Our framework was inspired by recent image inpainting approaches; it enables unsupervised motion completion, semantic animation modifications, as well as animation editing from sparse keyframes or coarse noisy animation signals.
The lack of high-quality animation data remains the major limitation in facial animation synthesis and editing research.
While our system obtains good results, we note that the quality of produced animation can only be as good looking and accurate as what the quality and diversity of our animation database covers.
We present various results, testifying for the validity of the proposed framework, but the current state of our result calls for experimentations on more sophisticated blendshape models, more diverse facial motions, and possibly the addition of rigid head motion.

The presented method relies on a generative model and offers no guarantee as such to match input constraints exactly. Yet, ensuring an exact hit is a standard requirement for high-quality production. We note that a workaround solution in our case would be to post-process our system's animation to match sparse constraints exactly, following the interpolation of ~\cite{seol_spacetime_2012} for instance.
Beyond the proposed solution for offline facial animation editing, an interesting direction would be to enable facial animation modifications to occur in real-time.
We plan on evaluating the performance of a forward-only recurrent network to assess the feasibility of real-time use cases.

Our system aims to make facial animation editing more accessible to non-expert users, but also more time-efficient, to reduce the bottleneck of animation cleaning and editing.
In terms of user interaction, our semantic editing framework requires isolating the animation segments to edit, and providing editing cues.
An interesting future work would be to integrate our system within a user-oriented application, combining our network with a user interface and a recording framework, forming a complete, interactive, efficient animation editing tool. 
Another interesting extension of this work would be to consider audio signals as additional input controllers.

\bibliographystyle{eg-alpha-doi}  
\bibliography{references}        



\end{document}